\documentclass[aps,prb, twocolumn,groupedaddress]{revtex4-1}
\newcommand{\up}{\uparrow}
\newcommand{\dw}{\downarrow}

\begin{document}

%\preprint{}
\title{Mechanism of Electron Spin Relaxation in Spiral Magnetic Structures}


\author{ I.I. Lyapilin}
\email {ligor47@mail.ru}
%\thanks{}
\affiliation{Institute of Metal Physics of Ural Branch of Russian Academy of Sciences, 620990 Yekaterinburg, Russia\\
Department of Theoretical Physics and Applied Mathematics, Ural Federal University, Mira St. 19, 620002 Yekaterinburg, Russia}


\date{\today}

\begin{abstract}
Spin dynamics in spiral magnetic structures has been investigated. It has been shown that the internal spatially dependent magnetic field in such structures produces a new mechanism of spin relaxation.

\end{abstract}
%\keywords{thermal perturbation; magnons; spin current; exchange interaction; thermoelectric coefficients}

\maketitle


\section{Introduction}

 One of the central problems facing spintronics, such as the development of methods of injection, generation, and detection of spin-polarized charge carriers, is to analyze and study the mechanisms of spin relaxation. Spin-orbit interaction (SOI) is well-known to have a significant influence on the mechanisms of electron spin relaxation. SOI can impact on spin degrees of freedom through translational ones \cite{1}. SOI itself does not cause spin relaxation but in combination with the scattering of the electron momentum leads to its relaxation. Elliott examined the process of electron spin relaxation through the momentum scattering at impurity centers, conditional upon inducing the spin-orbit interaction by lattice ions \cite{2}. Phonons can also participate in spin relaxation. The nature of spin relaxation involving phonons is to modulate time-dependent spin-orbit relaxation by lattice oscillations \cite{3}. As a result, spin relaxation occurs. The Elliott-Jafet mechanism offers for the spin relaxation frequency  $\omega_s$ to be proportional to the electron momentum relaxation frequency $\omega_p$.
   In systems without a center of inversion, a fundamentally different mechanism of spin relaxation can be realized. In such systems, the spin states $\up$ and  $\dw$ are not degenerate $E_{k\up}\neq E_{k\dw}$,  but the condition is satisfied $E_{k\up}= E_{-k\dw}$. The lack of a center of symmetry contributes to splitting the states subjected to SOI. The splitting can be described by entering an intrinsic magnetic field ${\bf B}_i({\bf k})$ around which the electron spins precess at a Larmor frequency. The precession of the electron spin in the effective magnetic field together with the electron momentum scattering leads to spin relaxation, and $\omega_s\sim \omega^{-1}_p$ \cite{4}.
   Below we consider the spin relaxation mechanism in magnetically ordered structures, which are characterized by a spiral arrangement of magnetic moments relative to some crystal axes \cite{5}. The simplest case of such structures is an antiferromagnetic spiral or a helicoid; they can be found in rare earth metals ({\it Eu, Tb, Dy}) and some oxide compounds. The structure of this type can be represented as a sequence of atomic planes perpendicular to the axis of a helicoid. In this case, atoms in each of the planes have ferromagnetically ordered magnetic moments. However, the magnetic moments in neighboring planes turn at some angle $\theta$ depending on the ratio of exchange interactions. This is because of the coexistence of positive exchange interaction between the nearest atomic neighbors and negative exchange interaction between the neighbors following the closest ones. The  components of the magnetic moments of atoms oscillate in the plane of the magnetic layer $S_x = S_{0x} sin kz,\,S_y = S_{0y} cos kz$. If, in this case, $S^z\neq 0$, we have a ferromagnetic spiral with a resulting moment. If   also oscillates $S^z$ by a harmonic law, a complex magnetic structure emerges. In such structures, the exchange interaction between zone charge carriers and localized moments is described by the well-known expression  $H_{ex} = - J\sum_i{\bf s}\cdot {\bf S}$ where $j$ is the exchange constant. In the mean-field approximation, this interaction can be represented as   $H_{ex} = -{\bf m}({\bf r}) H_{ef}({\bf r})$, where ${\bf m}({\bf r}) = g\mu_B {\bf S} ({\bf r}), g $ --   is the spectroscopic splitting factor, $\mu_B$  is the Bohr magneton, and $H_{ef}({\bf r})$  is the internal effective magnetic field.

   It is obvious that the spatial variation of the internal field in spiral magnetic structures should affect the spin dynamics of conduction electrons. Indeed, the spin of an electron in a state with quasi-momentum $({\bf k})$  precesses in the effective magnetic field $H_{ef}({\bf r})$  only for a time of the order of the elastic scattering time $\tau_p$ . After scattering, the electron goes into a state   $({\bf k}')$ where the effective magnetic field has a different direction. Consequently, the evolution of the spin dynamics under such conditions turns out to be associated with the electron momentum relaxation.

   Let us look into a simple spiral when the effective magnetic field   rotates in a plane $(xy)$, where $H_{ef} (H_0 cos (Qz),\, H_0 sin (Qz),\,0)$ , $ Q = 2\,\pi/\Lambda$, $\Lambda$  is the period of the spiral structure. To gain insight into the spin dynamics, we estimate the spin relaxation mechanism realized under above conditions. The system at hand is assumed to expose to an electric field directed along the axis z (${\bf E} = (0,\,0\,E_0)$ .
   The Hamiltonian of the system can be written in the form $   H = H_k + H_s + H_ {eE} + H_v + H_ {ev},$, where $H_k,\, H_s$ are the operators of the kinetic and spin energy of electrons interacting with the lattice $H_{ev}$  and the external electric field $H_{eE} = - e\sum_i {\bf E}\,{\bf r}_i$   and $H_v$  is the Hamiltonian of the lattice.
    \begin{eqnarray}\label{1}
    H_k = \sum_j p_i^2/2m,\qquad\qquad\qquad\,\nonumber\\ H_s =- g\mu_B\sum_j {\bf s}_i\, {\bf H}_{ef}= -\hbar\omega_{sf}\sum_j(s^+\,e^{iQz_j} + s^-_j\,e^{-iQz_j})\nonumber\\,
    \end{eqnarray}
where $\omega_{sf}= g e H_0/2m_oc$    is the electron precession frequency in the effective field, $s^\pm = s_x \pm i s_y$.
   The spin dynamics is determined by macroscopic equations of motion for the spin subsystem of electrons. The macroscopic equations of motion can be deduced by averaging microscopic equations of motion over the non-equilibrium statistical operator $\rho(t)$ . To begin with, we write down the microscopic equations of motion for the longitudinal components of the spin and momentum of electrons. We have
\begin{eqnarray}\label{2}
\dot{s}^z = i\omega_{sf}\{s^+\, e^{iQz}- s^-\,e^{-iQz}\} + \dot{s}^z_{sv},
 \end{eqnarray}
\begin{equation}\label{3}
\dot{p}^z =  -\,e\,E\,N+ iQ\hbar\omega_{sf}\{s^+\,e^{iQz}- s^-\,e^{-iQz}\}  +\dot{p}^z_{ev}({\bf r}),  .
\end{equation}
$\dot{A} =~(i\hbar)^{-1}[A,\,H],\qquad \dot{A}_{ij} =~(i\hbar)^{-1}[A,\,H_{ij}].$
Averaging the microscopic equations of motion over the operator $\rho(t)$, we arrive at  $<A_i>^t= Sp\{A_i\,\rho(t)\}$. The explicit form of the operator $\rho(t)$  can be found within the NSO method \cite{6}. Suppose that we have carried out this procedure. As a result, we come to the following system of macroscopic equations.
\begin{eqnarray}\label{4}
<\dot{s}^z>^t = i\omega_{sf}\{<s^+ e^{iQz}>^t - <s^-e^{-iQz}>^t\} + <\dot{s}^z_{ev}>^t,\nonumber\\
\end{eqnarray}
and
\begin{eqnarray}\label{5}
<\dot{p}^z> =  iQ\hbar\omega_{sf}\{<s^+\,e^{iQz}>^t - <s^-\,e^{-iQz}>^t\}-\nonumber\\
   -\,e\,E\,n_0 + <\dot{p}^z_{ev}>^t.\qquad\qquad\qquad
\end{eqnarray}
where $<N>^t=n$ is the electron concentration.
 The collisional summands in the balance equations (\ref{4}) and (\ref{5}) can be represented as follows \cite{6, 7}:
\begin{equation}\label{6}
<\dot{s}^z_{ev}>^t \sim (s^z,\,s^z)_0\,\omega_s,\quad <\dot{p}^z_{ev}>^t \sim\,(p^z,\,p^z)_0\,\omega_p,
\end{equation}
 where
 \begin{equation}\label{7}
(A,\,B)_0 = \int\limits_0^1 d\tau\,Sp\{A\rho_0^\tau\,\Delta B\,\rho_0^{1-\tau}\},
\end{equation}
 where $\Delta A = A -<A>_0,\, <\cdots>_0 = Sp\{\cdots \rho_0\}\, $, $\rho_0$ is the equilibrium Gibbs distribution.
 $\omega_s$   is the relaxation frequency of the longitudinal spin components, and  $\omega_p$ is the momentum relaxation frequency; both latter are calculated in the Born approximation over the scatter-electron interaction.
 \begin{eqnarray}\label{8}
\omega_s = \frac{g\mu_B}{\chi}\,\int\limits_{-\infty}^0\,dt \,e^{\epsilon t}\,(\dot{s}^z_{ev},\,\dot{s}^z_{ev}(t))_0,\nonumber\\
\omega_p = \frac{1}{m n T}\,\int\limits_{-\infty}^0\,dt \,e^{\epsilon t}\,(\dot{p}^z_{ev},\,\dot{p}^z_{ev}(t))_0,\quad
\epsilon\rightarrow +0 .
\end{eqnarray}
 Here $\chi$  is the paramagnetic susceptibility of an electron gas
\begin{equation}\label{9}
\chi =\frac{(g\mu_B)^2}{2 T}\cdot (s^+,\,s^-)_0 =\frac{g \mu_B}{H_0}\cdot<s^z>_0,
\end{equation}
$T\equiv k_B T$ is the temperature in energy units.
 Taking a stationary case into account, we finally obtain the expression for the spin relaxation frequency in a helicoidal magnet:
\begin{equation}\label{9}
\omega_s \sim \frac{m n T}{\hbar\,Q\,(s^z,\,s^z)_0}\cdot\omega_p
\end{equation}
   Thus, in spiral magnets as well as in the Elliott-Jafet mechanism, the spiral relaxation is related to the electron momentum relaxation and $\omega_s\sim \omega_p$. However, against the Elliott-Jafet mechanism, the spin relaxation mechanism in spiral magnetic structures is due to the presence of the internal spatially-dependent magnetic field in them. The evolution of the spin dynamics is determined by both the period of the magnetic structure and the momentum relaxation frequency.

   ACKNOWLEDGMENTS
The research was carried out within the state assignment of Minobrnauki of Russia
(theme  "Spin” No AAAA-A18-118020290104-2), supported in part by RFBR (project No. 19-02-00038/19).

\end{document}